\documentclass[twocolumn,twocolappendix]{aastex701}

\usepackage{amsmath}
\usepackage{comment}

\def\ergs{\rm{erg}\ \rm{s^{-1}}}
\def\Msunyr{\rm{M}_\odot\ \rm{yr}^{-1}}

\def\Msun{M$_\odot$}

\def\kms{km\,s$^{-1}$}

\newcommand{\NII}{{[N\,{\sc ii}]\ }}

\newcommand{\OIII}{{[O\,{\sc iii}]\,}}

\newcommand{\OII}{{[O\,{\sc ii}]\,}}

\newcommand{\Ha}{H$\alpha$\,}

\newcommand{\Hb}{H$\beta$\,}

\newcommand{\HeI}{{He\,{\sc i}\,}}
\def\kms{km\,s$^{-1}$}

\def\ergs{\rm{erg}\ \rm{s^{-1}}}

\shorttitle{Test LRD nature with radio and optical spectrosocpy}
\shortauthors{Mazzolari et al.}
\submitjournal{ApJL}

\begin{document}

\title{Radio emission from a local Little Red Dot: a challenge to the ionized cocoon scenario}

\correspondingauthor{Giovanni Mazzolari}
\author[orcid=0009-0005-7383-6655]{Giovanni Mazzolari}
\affiliation{Max-Planck-Institut f\"ur extraterrestrische Physik, Gie{\ss}enbachstra{\ss}e 1, 85748 Garching, Germany}
\affiliation{INAF -- Osservatorio di Astrofisica e Scienza dello Spazio di Bologna, Via Gobetti 93/3, I-40129 Bologna, Italy}
\email[show]{gmazzolari@mpe.mpg.de}

\author[orcid=0000-0003-4891-0794]{Hannah \"Ubler}
\affiliation{Max-Planck-Institut f\"ur extraterrestrische Physik, Gie{\ss}enbachstra{\ss}e 1, 85748 Garching, Germany}
\email{hannah@mpe.mpg.de}

\author{Grace E. V. Blomfield}  
\affiliation{Max-Planck-Institut f\"ur extraterrestrische Physik, Gie{\ss}enbachstra{\ss}e 1, 85748 Garching, Germany}
\email{gblomfield@mpe.mpg.de}

\author[orcid=0000-0002-5059-6848]{Marcella Brusa}
\affiliation{Dipartimento di Fisica e Astronomia ``Augusto Righi'', Universit\`a di Bologna, Via Gobetti 93/2, I-40129 Bologna, Italy}
\affiliation{INAF -- Osservatorio di Astrofisica e Scienza dello Spazio di Bologna, Via Gobetti 93/3, I-40129 Bologna, Italy}
\email{marcella.brusa3@unibo.it}

\author[orcid=0000-0003-3921-3313]{Jianhang Chen}
\affiliation{Max-Planck-Institut f\"ur extraterrestrische Physik, Gie{\ss}enbachstra{\ss}e 1, 85748 Garching, Germany}
\email{cjhastro@gmail.com}

\author[orcid=0000-0003-4949-7217]{Ric Davies}
\affiliation{Max-Planck-Institut f\"ur extraterrestrische Physik, Gie{\ss}enbachstra{\ss}e 1, 85748 Garching, Germany}
\email{davies@mpe.mpg.de}

\author[orcid=0000-0003-2388-8172]{Francesco D'Eugenio}
\affiliation{Kavli Institute for Cosmology, University of Cambridge, Madingley Road, Cambridge CB3 0HA, UK}
\affiliation{Cavendish Laboratory -- Astrophysics Group, University of Cambridge, 19 JJ Thomson Avenue, Cambridge CB3 0HE, UK}
\email{fd391@cam.ac.uk}

\author{Frank Eisenhauer}  
\affiliation{Max-Planck-Institut f\"ur extraterrestrische Physik, Gie{\ss}enbachstra{\ss}e 1, 85748 Garching, Germany}
\affiliation{Department of Physics, Technical University of Munich, 85748 Garching, Germany}
\email{eisenhau@mpe.mpg.de}

\author[orcid=0000-0002-2767-9653]{Reinhard Genzel}
\affiliation{Max-Planck-Institut f\"ur extraterrestrische Physik, Gie{\ss}enbachstra{\ss}e 1, 85748 Garching, Germany}
\affiliation{Departments of Physics and Astronomy, University of California, Berkeley, CA 94720, USA}
\email{genzel@mpe.mpg.de}

\author[orcid=0000-0001-8121-6177]{Roberto Gilli}
\affiliation{INAF -- Osservatorio di Astrofisica e Scienza dello Spazio di Bologna, Via Gobetti 93/3, I-40129 Bologna, Italy}
\email{roberto.gilli@inaf.it}

\author{Yannick Gilmer}  
\affiliation{Max-Planck-Institut f\"ur extraterrestrische Physik, Gie{\ss}enbachstra{\ss}e 1, 85748 Garching, Germany}
\affiliation{Universit\"ats-Sternwarte, Fakult\"at f\"ur Physik, Ludwig-Maximilians-Universit\"at M\"unchen, Scheinerstra{\ss}e 1, 81679 M\"unchen, Germany}
\email{y.gilmer@campus.lmu.de}

\author[orcid=0009-0009-1792-7199]{Shweta Jain}
\affiliation{Max-Planck-Institut f\"ur extraterrestrische Physik, Gie{\ss}enbachstra{\ss}e 1, 85748 Garching, Germany}
\email{shwetajain@mpe.mpg.de}

\author[orcid=0000-0002-1660-9502]{Xihan Ji}
\affiliation{Kavli Institute for Cosmology, University of Cambridge, Madingley Road, Cambridge CB3 0HA, UK}
\affiliation{Cavendish Laboratory -- Astrophysics Group, University of Cambridge, 19 JJ Thomson Avenue, Cambridge CB3 0HE, UK}
\affiliation{Institute for Astronomy, School of Physics, Zhejiang University, 310058 Hangzhou, China}
\email{jxhloading@gmail.com}

\author[orcid=0009-0003-7423-8660]{Ignas Juod\v{z}balis}
\affiliation{Max-Planck-Institut f\"ur Astrophysik, Karl-Schwarzschild-Stra{\ss}e 1, 85748 Garching, Germany}
\email{ijuodzb@mpa-garching.mpg.de}

\author[orcid=0000-0001-5470-305X]{Dharam V. Lal}
\affiliation{National Centre for Radio Astrophysics -- Tata Institute of Fundamental Research, Post Box 3, Ganeshkhind P.O., Pune 411007, India}
\email{dharam@ncra.tifr.res.in}

\author[orcid=0000-0001-6052-4234]{Xiaojing Lin}  
\affiliation{Department of Astronomy, Tsinghua University, Beijing 100084, China}
\affiliation{Steward Observatory, University of Arizona, 933 N Cherry Avenue, Tucson, AZ 85721, USA}
\email{xiaojinglin.astro@gmail.com}

\author[orcid=0000-0002-4985-3819]{Roberto Maiolino}
\affiliation{Kavli Institute for Cosmology, University of Cambridge, Madingley Road, Cambridge CB3 0HA, UK}
\affiliation{Cavendish Laboratory -- Astrophysics Group, University of Cambridge, 19 JJ Thomson Avenue, Cambridge CB3 0HE, UK}
\affiliation{Department of Physics and Astronomy, University College London, Gower Street, London WC1E 6BT, UK}
\email{rm665@cam.ac.uk}

\author[orcid=0000-0002-9087-2835]{Marco Mignoli}
\affiliation{INAF -- Osservatorio di Astrofisica e Scienza dello Spazio di Bologna, Via Gobetti 93/3, I-40129 Bologna, Italy}
\email{marco.mignoli@inaf.it}

\author[orcid=0009-0004-8724-9163]{Meghana Pannikkote}
\affiliation{Max-Planck-Institut f\"ur extraterrestrische Physik, Gie{\ss}enbachstra{\ss}e 1, 85748 Garching, Germany}
\email{meghana@mpe.mpg.de}

\author[orcid=0000-0002-7392-7814]{Eleonora Parlanti}
\affiliation{Scuola Normale Superiore, Piazza dei Cavalieri 7, I-56126 Pisa, Italy}
\email{eleonora.parlanti@sns.it}

\author[orcid=0000-0001-9680-7092]{Isabella Prandoni}
\affiliation{INAF -- Istituto di Radioastronomia, Via P. Gobetti 101, I-40129 Bologna, Italy}
\affiliation{Department of Astronomy, University of Cape Town, 7701 Rondebosch, Cape Town, South Africa}
\email{isabella.prandoni@inaf.it}

\author{Karla Rodriguez-Amber}  
\affiliation{Max-Planck-Institut f\"ur extraterrestrische Physik, Gie{\ss}enbachstra{\ss}e 1, 85748 Garching, Germany}
\email{karla@mpe.mpg.de}

\author[orcid=0000-0002-2125-4670]{Taro Shimizu}
\affiliation{Max-Planck-Institut f\"ur extraterrestrische Physik, Gie{\ss}enbachstra{\ss}e 1, 85748 Garching, Germany}
\email{shimizu@mpe.mpg.de}

\author[orcid=0000-0003-4226-7777]{Giulia Tozzi}
\affiliation{Max-Planck-Institut f\"ur extraterrestrische Physik, Gie{\ss}enbachstra{\ss}e 1, 85748 Garching, Germany}
\email{gtozzi@mpe.mpg.de}

\author[orcid=0009-0007-3295-8669]{Bartolomeo Trefoloni}
\affiliation{Scuola Normale Superiore, Piazza dei Cavalieri 7, I-56126 Pisa, Italy}
\email{bartolomeo.trefoloni@sns.it}

\begin{abstract}

Several scenarios have been proposed to explain the properties of Little Red Dots (LRDs), invoking either heavy obscuration by dense neutral gas, super-Eddington accretion, or an ionized gas cocoon that produces electron scattering. Radio observations can distinguish between these scenarios. Radio emission is largely unaffected by neutral gas and dust obscuration, whereas free-free absorption can suppress it in the presence of large free electron columns. Here, we investigate the origin of radio emission from the local LRD J1047. We jointly analyze its radio data and the broad H$\alpha$ emission from optical spectroscopy to test the predictions of different physical scenarios. We find that the observed radio luminosity must have an AGN origin, and it is consistent with that of a standard radio-quiet AGN obscured by dense neutral gas. Pure super-Eddington accretion underpredicts the radio luminosity by a factor of $\sim 30$, although a combination of super-Eddington accretion and obscuration remains viable. Interestingly, the statistically preferred model for the H$\alpha$ profile is an exponential fit, which is often associated with electron scattering in LRDs. This interpretation implies an electron column density of $N_e=2.4\times10^{24}\rm~cm^{-2}$ and a strong free-free absorption ($\tau_{\rm ff,5GHz}\sim 10^6$), incompatible with the radio detection. This tension suggests that the exponential profiles of J1047 more likely trace a stratified broad line region rather than electron scattering, with potential implications for virial black-hole mass estimates also in high-$z$ LRDs. Less likely, the radio emission may escape through a partially covering ionized medium or originate outside the putative scattering cocoon.

\end{abstract}



\section{Introduction}

One of the most exciting results of the first years of JWST operations is the discovery of Little Red Dots \citep[LRDs,][]{Matthee23}, a population of compact, broad-line AGN characterized by a distinctive V-shaped spectral energy distribution (SED) \citep[see][for a recent review]{Inayoshi26}. Many LRDs exhibit narrow absorption features in their broad Balmer lines \citep{Juodzbalis24_rosetta,Matthee26,Juodzbalis26_abs}, rare among AGN discovered before JWST \citep{Ji26_QSOabs,Hutchings02_narrowabs, Wang15_narrowabs}. Despite the evidence for accreting black holes (BHs) provided by their broad emission lines and luminosities, LRDs are remarkably X-ray weak, with stringent upper limits up to $\sim$2-3 dex below standard Type I AGN expectations \citep{Comastri26, Geris26,Yue24,Ananna24,Maiolino24_X,Mazzolari25_CEERS}.
Several scenarios have been proposed to explain these peculiar properties (see \citealt{Inayoshi26}).
In one scenario, dense neutral gas surrounding the central BH, with densities of $n\sim10^{9-10}\ \rm cm^{-3}$ and potentially high covering factors, produces the observed optical emission and Balmer features, while absorbing the intrinsic X-ray emission \citep{Inayoshi2024,Ji25,Juodzbalis25_AGNsample}. Alternatively, the X-ray weakness may be intrinsic, resulting from super-Eddington accretion, which can suppress and collimate the X-ray emission through the formation of a slim disk and a weak corona \citep{Pacucci24,Lambrides24,Madau2024_SE,King25,Tortosa26_Xweak}. In this scenario, the optical properties of LRDs can be reproduced when they are observed through the equatorial Broad Line Region (BLR) and dusty torus distribution \citep{Madau26_LRDLBD, Geris26, Lambrides24}.
Many LRDs also exhibit non-Gaussian broad-line profiles, well reproduced by exponential fits \citep{Rusakov25, Nikopoulos25, Deugenio25_irony}. These profiles have been interpreted as the result of Thomson scattering by free electrons in an ionized hydrogen layer \citep{Rusakov25,Torralba25,DeGraaff25_BB,Matthee26,Sneppen26_cocoon,Gentile26}. If so, these broad components would no longer trace the BLR dynamics, implying BH masses up to 1-2 dex below standard virial estimates and correspondingly higher accretion rates \citep{Rusakov25,Chang25,Chang26,Inayoshi26}.

Radio observations can discriminate among these scenarios \citep{Tanaka26_freefree,Mazzolari26_SKAO,Mazzolari26_radioJWST}. Unlike X-rays, synchrotron emission is generally unaffected by dust or neutral-gas obscuration \citep{Mazzolari24a,Ricci24} and can trace BH accretion through coronal activity, shocks, or small-scale jets, even in radio-quiet AGN \citep[see][for a review]{panessa19}. Therefore, standard AGN obscured by dense gas should remain detectable in the radio, whereas intrinsically X-ray-weak, super-Eddington systems are expected to produce also a fainter radio emission \citep[as shown in][]{Kellermann1989,Dong14, Paul26}. Conversely, the large free-electron columns invoked by the scattering scenario may produce strong free-free absorption (FFA), particularly at low frequencies, and alter the radio spectrum \citep{Walker00_freefree,Mazzolari26_SKAO,Rusakov25}.

All attempts to detect the radio emission of high-z LRDs failed \citep{Perger24, Gloudemans25,Mazzolari26_radioJWST}, with stacking analysis probing down to $f_{\rm 3GHz}\sim 0.2\mu$Jy (corresponding to $L_{\rm 3GHz}\sim 10^{39}\ergs$ at $z\sim 5$) and tentatively indicating a fainter radio emission than what is expected from standard Type I AGN. Recently, three local LRDs were found to have radio detections: J1047 \citep{Lin26_localLRD, Rodriguez26_radioLRD}, J0923 \citep{Lin25_LRDanalogs,An26_radioLRD}, J2048 \citep{Chen26_LRDanalogs,Rodriguez26b_radioLRD}. J2048 is embedded in a massive starburst galaxy (strongly contributing also to the radio output), J0923 has only shallow X-ray limits ($L_{\rm 2-10keV}<42.5$), and neither shows hydrogen or helium absorptions indicative of dense gas. Instead, J1047 is the source that most closely matches the properties of high-z LRDs and has the highest data quality, providing the best opportunity to use radio data to constrain the nature of LRDs.

Throughout this work, we assume a flat $\Lambda$CDM universe with $H_{0}=70 \; \rm{km s^{-1} Mpc^{-1}}$, $\Omega_{m}=0.3, \Omega_{\Lambda}=0.7$. We report typical synchrotron radio spectra in the form $S_{\nu}\propto \nu^{\alpha}$.

\section{Observations and results}
J1047 (RA: 161.98302, DEC: 7.66423), a galaxy at $z=0.1682$, was selected by \cite{Izotov07} as a compact, metal poor broad line emitter and studied by \cite{Simmonds16} and \cite{Burke21} as an example of an X-ray weak Seyfert galaxy, given its tentative 2-3$\sigma$ X-ray detection with {\it Chandra}, corresponding to $L_{\rm 2-10keV}=8^{+3.8}_{-4}\times 10^{40}\ergs $ \citep{Lin26_localLRD}. J1047 was classified as a local LRD in \cite{Lin26_localLRD}, which selected candidates in the SDSS DR17 using three high-$z$ LRD criteria: the V-shaped SED (matched to high-$z$ LRD templates), compact morphology (Petrosian radius $<1.8^{\prime\prime}$), and similar ionization conditions to high-$z$ LRDs (i.e. \OIII/\OII$>10$). With these criteria, all selected sources showed broad Balmer lines. J1047 features a broad \Ha component with $FWHM=952$ \kms and luminosity $L_{\rm H\alpha,broad}=2.75\times 10^{42}\ergs$, implying a BH mass of $\log (M_{\rm BH}/M_{\odot})=6.76$ using the virial calibration from \cite{Reines15}. Magellan/FIRE spectroscopy further revealed a deep absorption feature in the \HeI$\lambda 1.083\mu$m line. Multi-epoch photometry and spectroscopy (over 8 and 17 rest-frame years, respectively) show no variability above $3-4\%$ \citep{Burke26}, as observed also in high-z LRDs \citep{Secunda25}. Moreover, photoionization modeling of its broad Balmer and Paschen ratios indicates that they originate from dense, optically thick gas with $n_H\sim 10^{10-11}\rm \ cm^{-3}$ \citep{Kiyota26_LRDdecrement}.

\cite{Rodriguez26_radioLRD} reported J1047 radio detection from VLA observations in 2010 at 5~GHz ($f_{\rm 5GHz}=130\mu$Jy) and in 2018 at 6~GHz over a 4~GHz bandwidth ($f_{\rm 6GHz}=117\mu$Jy), corresponding to $L_{\rm 5GHz}=5\times 10^{38}\ergs$. The two measurements showed no variability over that period and yield $\alpha=-0.85\pm 0.24$, typical of optically thin synchrotron emission \citep{panessa19}.
Using the optical data in \cite{Lin26_localLRD}, we computed the radio-loudness parameter as in \cite{Kellermann1989}: $R=L_{\nu, 5GHz}/L_{\nu, 4400 A}=4.4$, placing J1047 in the radio-quiet AGN regime. 

We searched for radio counterparts of J1047 in the FIRST radio survey at 1.4~GHz \citep{White97_FIRST} and in the LOFAR DR3 \citep{Shimwell26_LOFARDR3}, finding none. The FIRST observations at the position of J1047 were performed in 2003, while the LOFAR mosaic at the source position (mosaic 1341) is dominated by observation P162+07 from 2024. The measured local rms values are $125\,\mu$Jy at 144 MHz and $150\,\mu$Jy at 1.4 GHz. While the FIRST non-detection is consistent with the high-frequency spectral index, which predicts $f_{\rm 1.4GHz}=380\,\mu$Jy ($<3\times \rm rms_{FIRST}$), the LOFAR non-detection is unexpected, as we discuss in Sect.~\ref{sec:LOFAR}.
Combining J1047's radio and spectroscopic properties allows us to test scenarios proposed to explain the nature of LRDs and assess their implications for the broader LRD population.

\subsection{LRD properties from \Ha fitting}\label{sec:Hafit}
\begin{figure*}[ht]
    \centering
    \includegraphics[width=0.329\linewidth]{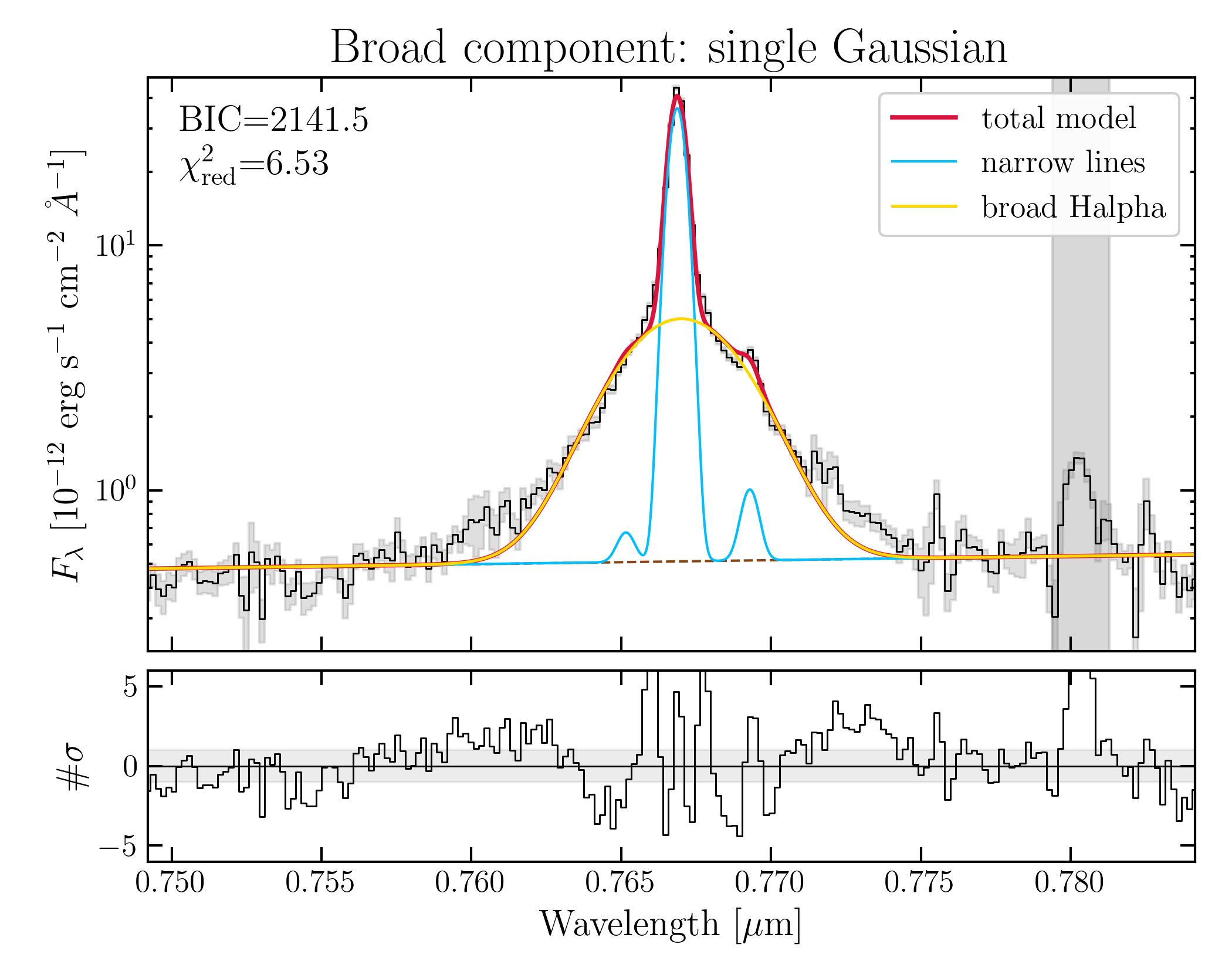}
    \includegraphics[width=0.329\linewidth]{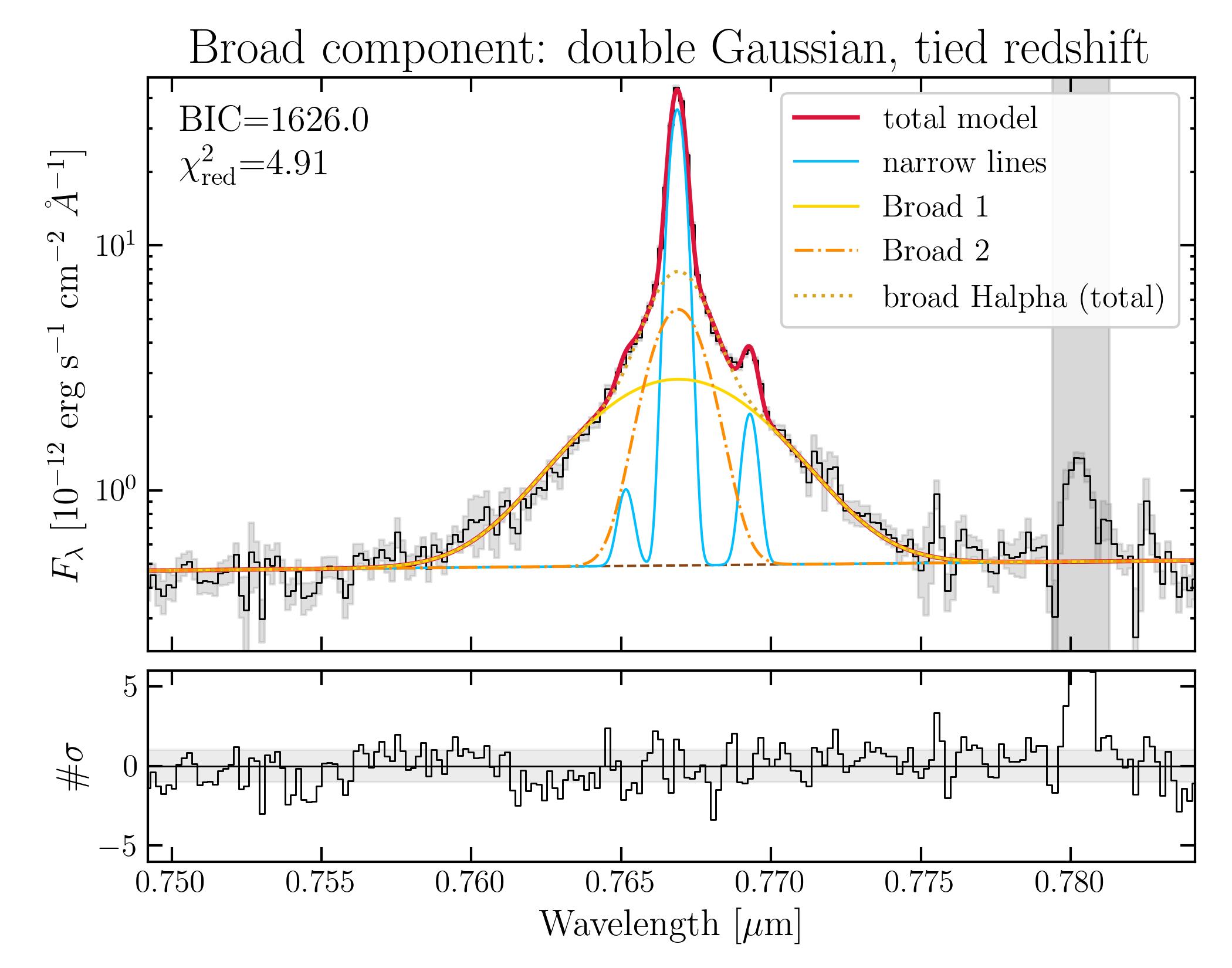}
    \includegraphics[width=0.329\linewidth]{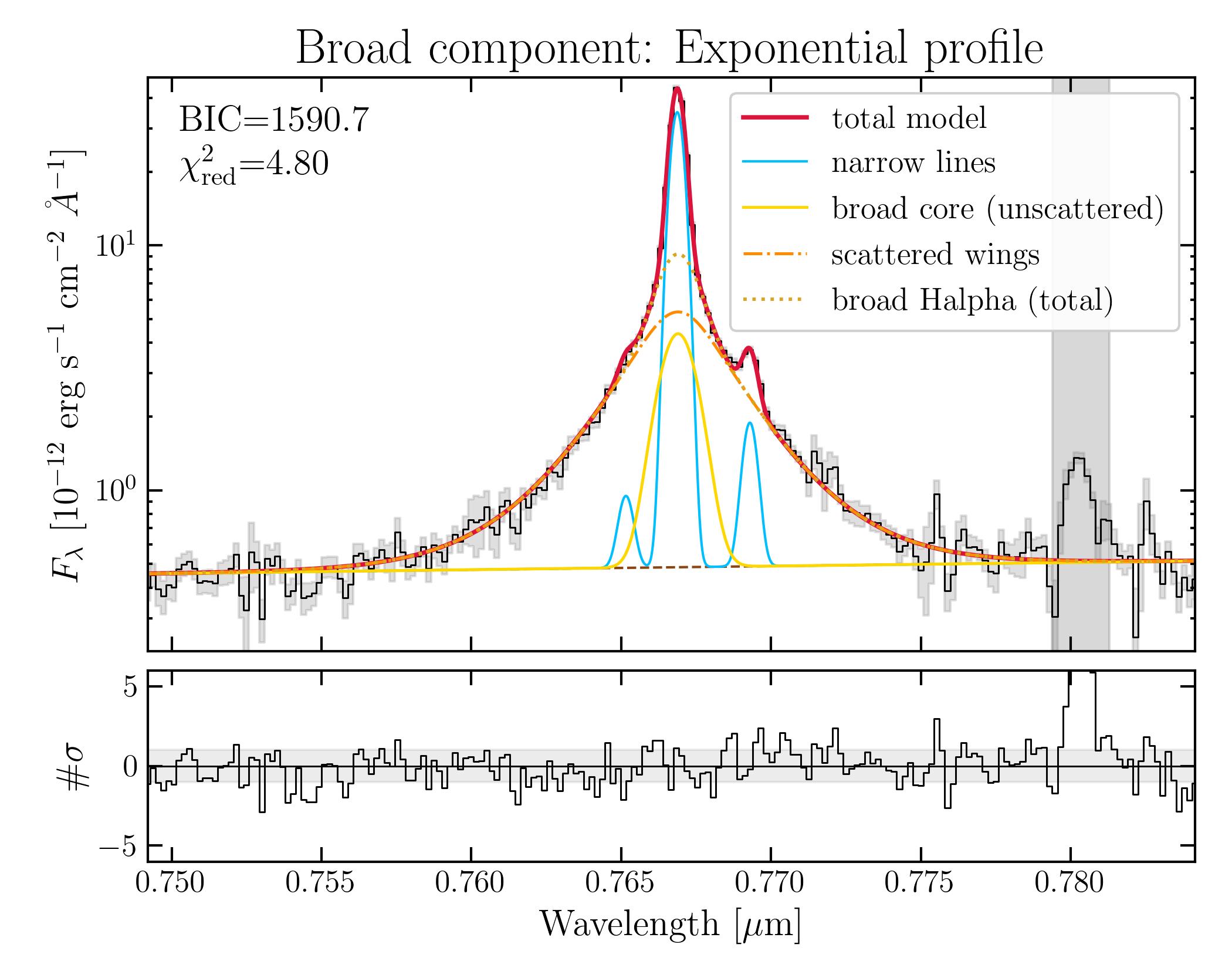}
\caption{Fits to the broad \Ha line profile of J1047, as described in Sect.~\ref{sec:Hafit}. Left: single Gaussian; centre: double Gaussian with tied redshifts; right: exponential profile. BIC and $\chi_{red}^2$ are indicated in each panel, computed as described in Appendix~\ref{sec:app_Hafit}. The gray shaded region marks the position of the \HeI$\lambda 6680\AA$ line, not included in the fit. The bottom panels show the fit residuals. }
    \label{fig:line_profiles}
\end{figure*}
J1047 was observed spectroscopically in 2003 (SDSS), 2016 (Keck), and 2020 (Gemini). We jointly analyze data obtained at similar epochs, to minimize effects of variability that could have impacted the data before or afterwards, as observed in the radio band for the other radio-detected LRD J0923 \citep{An26_radioLRD}. However, the negligible variability among the three optical spectra \citep{Burke26} makes the choice largely irrelevant over the VLA epochs, while no optical follow-up is simultaneous with the LOFAR observation.
Most high-z LRDs are characterized by non-Gaussian broad Balmer lines, and frequently better described by exponential profiles \citep{Rusakov25,Scholtz26_expo}. These exponential profiles have often been interpreted as evidence of electron (Thomson) scattering broadening the lines \citep{Rusakov25, DeGraaff25_BB, Torralba25, Chang25, Chang26}. We test this possibility for J1047 by fitting its SDSS spectrum around \Ha and \NII with three different models (see Appendix~\ref{sec:app_Hafit}).
All models include narrow \Ha and \NII components, while the broad \Ha component is described by either a single Gaussian, two Gaussians with independent widths and amplitudes but tied redshifts, or as a Gaussian component convolved with a symmetric exponential following \cite{Laor06}: $E(\lambda_0,W; \lambda) \propto e^{-\frac{|\lambda-\lambda_0|}{W}}, $ where $\lambda_0$ is the central wavelength (the same of the Gaussian), and $W$ is the exponential width. The first two models provide a simplified description of the broad lines that is often assumed for high-z AGN \citep{Scholtz26_expo}, while in the third the Gaussian represents the intrinsic BLR emission, but with a fraction of it Thomson-scattered by free electrons. In the latter, the full profile of the broad line is therefore given by: 
\begin{equation}
    f_{scatt} E(\lambda) * G_{BLR}(\lambda) + (1 - f_{scatt})G_{BLR}(\lambda),
\end{equation} 
where $f_{scatt}$ is the fraction of scattered light \citep{Scholtz26_expo}.

The three fits are reported in Fig~\ref{fig:line_profiles}.
The exponential model is strongly favored over the single- and double-Gaussian models ($\Delta BIC=-550$ and $-35$, respectively; see Appendix~\ref{sec:app_Hafit}). According to the interpretation of \cite{Rusakov25, DeGraaff25_BB}, this would imply that electron scattering is the dominant line-broadening mechanism, and the best-fit parameters can constrain the scattering medium's electron column density and optical depth. Following \cite{Scholtz26_expo}, the scattering fraction is related to these quantities by:
\begin{equation}
    \tau_{e}=-\ln(1-f_{\rm scatt}); \qquad N_e=\tau_{e}/\sigma_{T},
\label{eq:Ne}
\end{equation}
where $\sigma_T=6.65\times10^{-25} \rm \ cm^{2}$ is the Thomson cross section. The fit returned a scattering fraction of $f_{scatt}=80\pm1 \%$, corresponding to $N_e=2.4\times 10^{24}\rm \ cm^{-2}$.  In addition, following the approximation of \citet{Rusakov25}, the exponential width $W$ is related to the electron temperature by:
\begin{equation}
W = (428 \times \tau_{thom} + 370) \times \biggl(\frac{T_e}{10^4 K} \biggr)^{0.5}\rm \ km\ s^{-1}. 
\label{eq:Te}
\end{equation}
We found $W=766^{+21}_{-18}$\kms, implying $T_e=5273$K.

\subsection{LRD properties from radio emission}\label{sec:radio_scenarios}
We first focus on the VLA high-frequency data, which were obtained during a period when no variability was observed in either the radio or optical. We then discuss the implications of the low-frequency LOFAR non-detection.
We compare the observed radio luminosity of J1047 with the predictions from three scenarios proposed to explain the properties of LRDs: (1) standard Type I AGN radio emission, with X-ray obscuration and the optical features explained by the presence of dense neutral gas and dust; (2) super-Eddington accretion, intrinsically producing weak X-rays; and (3) AGN fully surrounded by a dense, ionized scattering medium. We will estimate the radio luminosity predicted by each scenario using the physical quantities consistent with its underlying assumptions and then assess their consistency with the observed radio data and the results of the H$\alpha$ fitting.

\subsubsection{Standard AGN with dense gas obscuration}\label{sec:standardAGN}
To constrain the expected radio luminosity in the first scenario, we assume the calibrations valid for standard Type I AGN. In particular, we derive the expected radio luminosity using the fundamental plane correlations, a family of relations in the form: $\log L_R=\xi_{RX}\,\log L_X + \xi_{RM} \log M_{\rm BH}$ \citep{Merloni02,merloni03}. These correlations hold over several orders of magnitude in $M_{\rm BH}$ and $L_{\rm 2-10keV}$, although their coefficients depend on the BH accretion state \citep{Li08,Dong14,dong21,Wang24_FP,Bariuan22}. We adopt the relation derived by \cite{Li08} for radio-quiet Type I AGN:
\begin{equation}
\begin{split}
&\log \frac{L_{\mathrm{1.4~GHz}}}{10^{40} \ergs} =\\
&0.73\,\log  \frac{L_{\mathrm{0.1-2.4keV}}}{10^{44} \ergs} + 0.31\,\log \frac{M_{\mathrm{BH}}}{10^{8} \ M_{\odot}} -0.68,
\label{eq:fund_plane}
\end{split}
\end{equation}

that has a $\sigma=0.42$dex scatter.
Unlike most fundamental-plane calibrations, this relation is based on an optically selected sample of local Type I AGN (as in J1047), without pre-selection in accretion state, and also includes Narrow-line Seyfert 1 (NLS1) galaxies, which have spectroscopic properties similar to LRDs \citep{Berton25,Maiolino24_X,Deugenio26_pol,Geris26}.
In this scenario, we adopt $\log(M_{\rm BH}/M_\odot)=6.76$ based on the virial calibration of \cite{Reines15}, and assume that the observed X-ray luminosity is suppressed by dense gas and therefore does not represent the intrinsic luminosity required by Eq.~\ref{eq:fund_plane}.
Using the Type I AGN $L_{\rm 2-10keV}$--$L_{\rm H\alpha,broad}$ relation of \cite{Jin12}, we infer an intrinsic $\log(L_{\rm 2-10keV}/\ergs)=43.47$, 2.6 dex above the observed value. We convert this to the rest-frame $0.1$--$2.4$ keV band assuming an intrinsic photon index $\Gamma=1.8$ and obtaining $\log(L_{\rm 0.1-2.4keV}/\ergs)=43.58$. Unlike the X-rays, the radio emission remains largely optically thin and unaffected by obscuration. This scenario predicts $L_{\rm 5GHz}=5.2\times 10^{38}\ergs$, consistent with the observations.
\begin{figure*}[ht]
    \centering
    \includegraphics[width=1\linewidth]{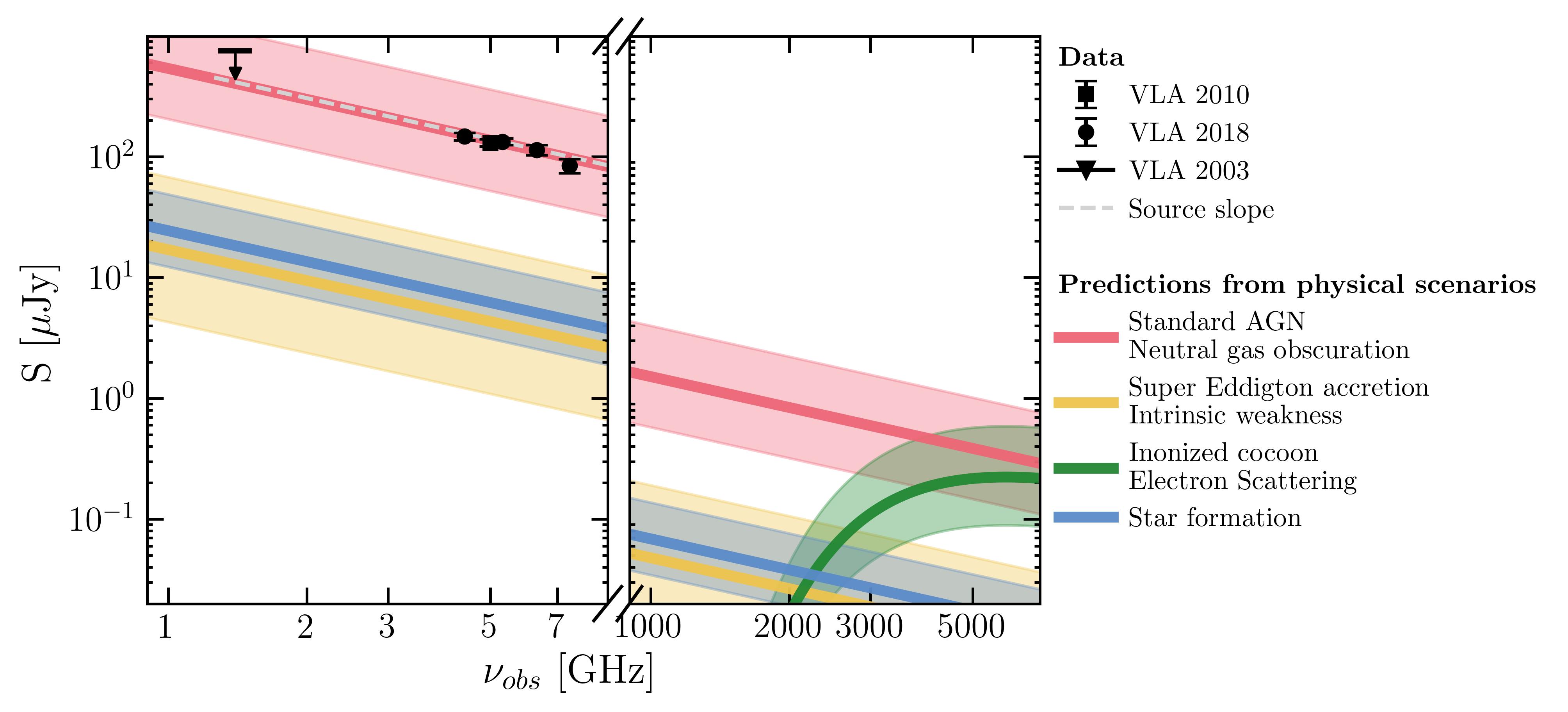}
    
    \caption{Radio detection of local LRD J1047 (black data points) and best fit to the observed radio spectral slope (dashed gray line), compared to the radio flux predictions from the models described in Sect.~\ref{sec:radio_scenarios}. The 2018 VLA data, collected over a 4~GHz bandwidth around 6~GHz, are split into 4 different channels, as reported in \cite{Rodriguez26_radioLRD}. The x-axis is cut to also show the prediction from electron scattering, which otherwise fully suppresses the radio emission at the frequencies of the J1047 detections.}
    \label{fig:spectral_slope}
\end{figure*}
\subsubsection{Super Eddington accretion and intrinsic weakness}
In the second scenario, we assume that the source is accreting at super-Eddington rates, which results in intrinsically weak X-ray emission, as shown by \cite{Pacucci24, King25, Madau2024_SE, Tortosa26_Xweak}, and which can also reproduce the optical properties of LRDs \citep{Kaaz26_LRDSE,Madau26_LRDLBD, Zucchi26_highion, Lambrides24}. An intrinsic X-ray weakness of J1047 is suggested also by its similarity to NLS1 which are characterized by broad components of $\sim 1000$\kms, high accretion rates, and a fainter and softer X-ray emission ($\sim 3-5$ times fainter than standard AGN \citealt{Berton18,Vaughan99, Maiolino24_X}). In this case, we estimate the expected radio luminosity using the fundamental plane correlation defined by \cite{Paul26} for super-Eddington accreting local AGN:
\begin{equation}
\log L_{\mathrm{5~GHz}} = 0.39\,\log L_{\mathrm{2-10keV}} + 0.39\,\log M_{\mathrm{BH}} + 18.63,
\label{eq:fund_plane2}
\end{equation}
which has a scatter of $\sim0.6$ dex.
In this scenario, assuming that the observed X-ray detection traces the intrinsic X-ray emission, and adopting the virial BH mass, we obtain $L_{\rm 5GHz}=1.5\times 10^{37}\ergs$. Recently, \cite{Madau26_LRDLBD} and \cite{Sneppen26_Xweak} proposed that the LRD X-ray weakness might come from a combination of super-Eddington accretion and gas obscuration. To have a radio luminosity from Eq.~\ref{eq:fund_plane2} consistent at 1$\sigma$ with the observed one, the $L_{\rm 2-10keV}$ needs to be corrected by a factor $140$ ($\sim 2.1$ dex), more typical for Compton-thick obscuration \citep{Levenson06_CTK, gilli11, Koss15_NGC3393} and alleviating the need to explain the whole observed X-ray suppression (2.6 dex) through absorption alone.

\subsubsection{Ionized cocoon and free-free absorption}\label{sec:radio_freefree}
The third scenario invokes a high-covering-factor ionized hydrogen layer that produces Thomson-scattered broad lines, leading also to a significant overestimate of BH mass in LRDs \citep{Rusakov25,Chang25,Chang26}. This layer may lie inside the neutral cocoon responsible for X-ray obscuration and optical emission \citep{DeGraaff25_BB}, or at slightly larger radii \citep{Torralba25}. In either configuration, its free-electron column may absorb the central radio emission through FFA \citep{Odea98,Laor08,Baskin21}. We conservatively assume an intrinsic radio luminosity typical of radio-quiet AGN as estimated with Eq.~\ref{eq:fund_plane}, but accounting for two additional effects. First, because scattering broadens the observed \Ha profile, we use only its intrinsic, unscattered component to estimate the virial BH mass. Adopting $FWHM_{\rm broad,int}=567$ \kms and the relation of \cite{Reines15}, we obtain $\log(M_{\rm BH}/M_\odot)=6.26$. Second, we account for FFA, which attenuates the radio luminosity as $L_R^{\rm obs}=L_R^{\rm int}e^{-\tau_{\rm ff}}$, where:
\begin{equation}\label{eq:tauff_EM}
    \rm \tau_{ff}=3.28\  10^{-7}\times \biggl(\frac{T_e}{10^4 K}\biggr)^{-1.35}
    \biggl(\frac{\nu_{rest}}{GHz}\biggr)^{-2.1}
    \biggl(\frac{EM}{pc\ cm^{-6}}\biggr),
\end{equation}
Here, $T_e$ is the electron temperature, $\nu_{\rm rest}=(1+z)\nu_{\rm obs}$, and $EM=\int n_e^2\,dl$ is the emission measure. For a uniform ionized slab of thickness $r$, $EM=N_e^2/r$, where $N_e$ is the electron column density; thus, Eq.~\ref{eq:tauff_EM} becomes:
\begin{equation}\label{eq:tau_ff_Ne}
   \rm  \tau_{ff}=3 \times \biggl(\frac{N_e}{10^{22} cm^{-2}}\biggr)^2
    \biggl(\frac{T_e}{10^4 K}\biggr)^{-1.35}
    \biggl(\frac{\nu_{rest}}{GHz}\biggr)^{-2.1}
    \biggl(\frac{r}{1 pc}\biggr)^{-1}.
\end{equation}
To reproduce the \Ha exponential wings, the ionized layer must have the $N_e$ and $T_e$ inferred from the line fit (Eqs.~\ref{eq:Ne} and \ref{eq:Te}). For a slab thickness of $r=0.01$ pc, these values imply $\tau_{\rm ff}\sim1.1\times10^6$ at 5 GHz, completely suppressing the radio emission. In Appendix~\ref{app:tauff_grid} we show that similar suppression of the radio emission would come also from other configurations of the scattering medium as derived in \cite{Wada26} for a local Type I AGN and in \cite{Torralba25} for an high-$z$ LRD.

\subsubsection{Star Formation}\label{sec:SFR}
Finally, we also consider the possibility that radio emission is produced by star formation.  
If all the radio emission of J1047 were attributed to star formation, it would imply a star formation rate (SFR) of $\sim 20~\Msunyr$, using the conversion from \cite{Murphy11,Kennicutt12}.
We first estimate the SFR from SED fitting, which is challenging for LRDs given the limited power of standard AGN templates and models to reproduce their shape \citep{Mazzolari26_NOEMA}. Using CIGALE (details reported in Appendix~\ref{sec:app_sed}) the best fit solution attributes the UV and a substantial optical contribution to the host galaxy emission ($\sim 60\%$ at $\lambda_{rest}=6500\AA$), contrary to the findings of \cite{Lin26_localLRD}, who found an AGN contribution to the rest-frame optical of $\sim85\%$. The resulting 100-Myr-averaged SFR (the timescale probed by radio emission) is $SFR_{\rm 100Myrs}=2.83\pm0.65\,\Msunyr$, potentially contributing only $\sim 15\%$ to the total radio luminosity, and the stellar mass $M_\star=1.5\times10^9$\Msun. Fixing the rest-optical AGN contribution to be as in \cite{Lin26_localLRD} the SFR drops to $SFR_{\rm 100Myrs}=1.35\pm0.15\,\Msunyr$.

As an independent estimate, we infer the SFR from the dynamical mass and assuming J1047 is on the main sequence. We fit the SDSS \OIII narrow line component and obtained a line-spread-function-corrected $FWHM_{\rm [OIII],narrow}\sim140$ \kms (consistent with the intrinsic narrow \Ha). Combining this dispersion with the HST effective radius ($R_e=568$ pc; \citealt{Lin26_localLRD}) in $M_{\rm dyn}\simeq5\sigma^2R_e/G$, we obtain $\log(M_{\rm dyn}/M_\odot)=9.46$. Conservatively assuming the dynamical mass to be dominated by stars, the main sequence of \cite{Popesso23} gives $SFR_{\rm MS}=0.9\ \Msunyr$, consistent with the SFR estimate in the AGN dominated fit. Even accounting for the 0.3 dex scatter of the \cite{Popesso23} main sequence, the SFR is too low to justify the radio luminosity. We assume this as our fiducial value. 

As a third estimate, we use the rest-frame UV emission traced by the GALEX FUV and NUV photometry, which gives a steep slope of $\beta_{UV}=-2.63$, bluer than the typical value for local star-forming galaxies ($\beta_{\rm UV}\sim-2.3$; \citealt{Reddy18a}) and suggesting negligible attenuation. Using the UV--SFR calibration of \cite{Madau14}, corrected for a Chabrier IMF ($f_{\rm IMF}=0.66$), we obtain $SFR_{\rm UV}\sim0.91\,\Msunyr$, consistent with the previous estimates.

If entirely ascribed to star formation, the \Ha luminosity of $L_{\rm H\alpha,narrow}=1.5\times10^{42}\ergs$ instead yields $SFR_{\rm H\alpha}=8.5\,\Msunyr$ using \cite{Kennicutt12}, potentially accounting for almost half the radio emission. However, this is an upper limit because narrow \Ha is likely contaminated by AGN photoionization. Additionally, it traces star formation over $\sim10$ Myr rather than the $\sim100$ Myr probed by radio emission. Moreover, this SFR would predict $f_{\rm 144MHz}=1.2$ mJy (always assuming $\alpha=-0.85$), inconsistent with the LOFAR non-detection; especially given that the host galaxy's radio emission is less likely to exhibit short-timescale variability or FFA.

\subsubsection{LOFAR non detection}\label{sec:LOFAR}
All the relations reported in the previous scenarios are calibrated on high-frequency radio data, where J1047 shows a quite steep spectral slope. Extrapolating this slope to LOFAR frequencies predicts $f_{\rm 144MHz}=2.65$ mJy, about four times the $5\sigma$ upper limit of $625\,\mu$Jy. The discrepancy may have different explanations, given the large frequency separation and six-year gap between the LOFAR and last VLA observations.
First, the intrinsic radio spectrum may flatten or curve below the VLA frequencies. A single power law consistent with the LOFAR limit requires $\alpha\gtrsim-0.4$, about $2\sigma$ flatter than measured. Forcing the fit (with 3 degrees of freedom) to remain below the LOFAR limit increases the $\chi^2$ of the five VLA data points from 1.7 to 5.8 ($\Delta\chi^2\simeq4.1$). Although this is disfavoured at approximately the $2\sigma$ level, it remains statistically acceptable ($p\simeq0.12$), and a flatter spectrum therefore cannot be excluded.

Alternatively, the LOFAR non detection could be due to variability, as observed in another radio-detected LRD \citep[J0923][even if at high-frequencies]{An26_radioLRD}. This would require a decrease by a factor of at least four at 144 MHz, which is unusual for low-frequency emission \citep{Hajela19,Ross21} and no spectroscopic data from 2024 are available to test this.

A third possibility is a low-frequency spectral turnover caused by synchrotron self-absorption. In this scenario, a compact synchrotron-emitting plasma becomes optically thick at low frequencies, suppressing the observed flux below the turnover frequency, which is connected to the size of the emitting region \citep{Ighina24}. A peak frequency of $0.3\lesssim\nu_{\rm p}/{\rm GHz}\lesssim3$ would be sufficient to reproduce the LOFAR non-detection while remaining consistent with the VLA measurements. However, the lack of data between 144 MHz and a few GHz prevents us from more precisely locating the peak frequency and testing this scenario further. 

Finally, FFA by a modest ionized gas column along the line of sight could suppress the 144 MHz emission while leaving the 4–7 GHz flux largely unaffected. As we show in Appendix~\ref{app:LOFAR}, a negligible suppression at $\sim 3$GHz and by a factor of at least 4 at 144 MHz requires $\tau_{e}\approx7\times 10^{-5}-7\times10^{-4}$, corresponding to of ${\rm N_e}\approx10^{20}-10^{21}{\rm cm^{-2}}$ for $T_e=10^4$K. Crucially, this column is far below that required for the electron-scattering interpretation of the \Ha profile. While modest FFA remains only one of the possible explanation for the LOFAR non-detection, it cannot be associated with the dense ionized medium required to produce the broad wings in the \Ha line.

\section{Discussion \& Conclusion}\label{sec:discussion}
As shown in Fig.~\ref{fig:spectral_slope}, the observed radio emission of J1047 is very well reproduced by that of a standard Type I AGN obscured by dense neutral gas suppressing the X-rays and producing the rest-frame optical features of LRDs, while remaining optically thin at radio waves. A purely super-Eddington scenario underpredicts the observed radio emission by a factor $\sim 30$. However, a combination of super-Eddington accretion and obscuration may still be consistent with the data: an intrinsic X-ray weakness by a factor of $3-5$ alleviates the need for extreme obscuration, bringing the required X-ray suppression closer to that typically expected for Compton-thick sources ($\sim 1\%$). Instead, the electron column and temperatures inferred from the exponential \Ha fit produce strong FFA, fully suppressing the high-frequency radio emission, while we showed that smaller columns potentially explaining the LOFAR non detection cannot explain the \Ha profile either. 
Additionally, the radio emission of J1047 must have an AGN origin, since the host galaxy SFR is $\sim 7-20\times$ lower than needed to reproduce the observed data.

The statistically favoured \Ha exponential fit is therefore difficult to reconcile with the radio detection if interpreted as a consequence of electron scattering.
However, while electron scattering naturally produces exponential line profiles, an exponential profile does not uniquely imply electron scattering \citep{Laor06,Kollatschny13_BLRprofiles, Peterson05_NGC4395, Scholtz26_expo, Madau26_wings, Brazzini25}, and non-Gaussian line profiles are common among Type I AGN \citep[][B.~Trefoloni et al.~in prep.]{Kollatschny13_BLRprofiles,Kocevski26}.
An alternative interpretation is that the exponential profile observed in J1047, and in other high-S/N high-$z$ LRD spectra, arise from a superposition of BLR clouds spanning a range of radii and thus virial velocities, rather than from scattering \citep{Feng25,Scholtz26_wings,Madau26_wings}. The BLR stratification preserves the BH mass estimated based on virial relations (even if with some corrections; \citealt{Scholtz26_expo}), contrary to the $\gtrsim 1$dex negative correction implied by electron scattering \citep{Rusakov25,Juodzbalis25_Mdyn}. The BLR stratification hypothesis is consistent with the assumptions used to predict the radio emission from the `standard' radio-quiet AGN in Sect.~\ref{sec:standardAGN}, which provides the best description of the radio data. These results, suggesting that the gas cocoon scattering scenario is incompatible with the radio measurement for J1047 may have implications also for the ongoing debate on BH mass estimates in LRDs and, more broadly, in high-z AGN \citep{Ubler26_BHreview}.

Alternatively, to reconcile the scattering scenario with the radio detection, the ionized gas `cocoon' may only partially cover the central source, with holes allowing the radio emission to escape. In this case, it would resemble a high-covering-factor BLR rather than a fully enclosing layer. While some studies required an almost $4\pi$ covering to explain LRD properties \citep{Gentile26,Naidu25}, others invoked partial covering and orientation effects to explain the differences between LRDs and bluer X-ray weak Type I AGN discovered by JWST \citep{Ji26_Lya,Tang26_Lya,Geris26,Matthee26,Sneppen26_LBD}.
Nevertheless, it remains difficult to explain how radio emission escapes along the same line of sight where the \Ha profile is dominated by scattering. A clumpy cocoon could provide escape channels \citep{Deugenio26_pol,Ji26_Lya,Tang26_Lya,Chang26}, but clump motion might be expected to produce spectral or photometric variability, which is not observed in J1047 over several rest-frame years.

A third possibility is that the radio emission originates from outside the ionized cocoon. The unresolved radio observations of J1047 constrain the emitting region to $<700$ pc, larger than the optical effective radius ($R_e=568$ pc), and cannot establish the actual scale of radio emission. We can estimate the minimal scale of the radio-emitting region using Eq.~22 of \cite{Laor08}:
\begin{equation}\label{eq:Rrad}
\begin{split}
    &\mathrm{\frac{R_{\rm rad}}{pc}=}\\
    &\mathrm{0.47\biggl(\frac{L_{\nu,rest}}{10^{30} erg\, s^{-1}\, Hz^{-1}}\biggr)^{0.4} 
    \biggl(\frac{L_{bol}}{10^{46} erg\, s^{-1} }\biggr) ^{0.1}
    \biggl(\frac{\nu_{rest}}{GHz}\biggr)^{-1},}
\end{split}
\end{equation}
where $L_{\rm \nu,rest}$ is the radio luminosity at rest-frame frequency $\nu_{\rm rest}$, and $L_{\rm bol}$ is the bolometric luminosity. Adopting the LRD-specific bolometric calibration $L_{\rm bol}=9L_{\rm H\alpha,broad}$ from \cite{Greene25}, we obtain $R_{\rm rad}\sim0.015$ pc. A $\sim10\times$ larger bolometric correction, as in \citep{Stern12}, increases this by only $\sim20\%$. Since Eq.~\ref{eq:Rrad} predicts smaller emitting regions at higher frequencies, extrapolating the radio flux to e.g.~20 GHz gives $R_{\rm rad}\sim0.003$ pc. The BLR radius, which is expected to trace also the location of the scattering gas \citep{Juodzbalis24_rosetta,Torralba25}, is $R_{\rm BLR}\sim0.025$ pc, using the $R_{\rm BLR}$--$\lambda L_{5100\AA}$ relation by \cite{GRAVITY24_BLRsize} and the observed $\lambda L_{5100\AA}=1.8\times10^{43}\ergs$. This is about 2 and 10 times larger than the minimum radio-emitting sizes at 5 and 20 GHz, respectively. Thus, the radio emission can originate from within the putative scattering region (but not necessarily, since Eq.~\ref{eq:Rrad} defines only the minimum scale). However, if radio originates from outside the cocoon (and the BLR) it would rule out most of the processes expected to produce radio emission in radio-quiet AGN, such as low-power jets or a magnetically heated corona (happening close to the BH \citealt{panessa19}). Indeed, \cite{Chen23_RQ,Chen26_RQ}, using high-resolution VLBA observations of a sample of radio-quiet AGN, showed that their radio emission comes from a scale very close to the BH, at approximately half the size of the BLR. If the low-frequency flattening of J1047 is due to synchrotron self-absorption (see Sect.~\ref{sec:LOFAR}), future observations between 144 MHz and 3 GHz could constrain the source size through the detection of the turnover frequency.

Our results demonstrate the diagnostic potential of radio observations in constraining the physical conditions of LRDs and highlight the importance of combining radio and optical spectroscopic diagnostics to distinguish between competing interpretations and test their implications for BH mass estimates. Further spatially resolved, multi-frequency radio observations will be crucial for determining the location and physical origin of the radio emission from J1047, while larger samples of radio-detected LRDs will be essential to perform the same analysis on a statistical population, ideally both at low and high-$z$. 

No high-$z$ LRD radio detection has been reported so far, and at the mean redshift of the JWST AGN sample studied in \cite{Mazzolari26_radioJWST} ($z\sim5.2$), the radio luminosity of J1047 would correspond to a flux density of $f_{\rm 5GHz}=0.05,\mu$Jy, approximately a factor of four below the rms reached by their stacking analysis. Thus, even a source with radio properties similar to J1047 would remain undetected at the sensitivity of current observations, and the existing high-redshift radio non-detections do not rule out a population with similar radio properties.
Future deep, multi-frequency continuum surveys with the SKA Observatory (SKAO), reaching $\rm rms\sim 0.05-0.1\mu$Jy with 100-1000 hours at $1.4-3$~GHz, are expected to eventually detect the emission of high-$z$ LRDs \citep{Mazzolari26_SKAO}. These will be essential for determining whether the radio properties of J1047 are representative of the broader LRD population and for distinguishing among the physical scenarios considered in this work. 

\begin{acknowledgments}
GM and H\"U acknowledge funding by the European Union (ERC APEX, 101164796). Views and opinions expressed are those of the authors only and do not necessarily reflect those of the European Union or the European Research Council Executive Agency. Neither the European Union nor the granting authority can be held responsible for them.
H\"U acknowledges support by the Max Planck Society through the Lise Meitner Excellence Program. 
DVL acknowledge support from the Department of Atomic Energy, under project, ``Next Generation Instrumentation for Radio Astronomy'' and the project UID number is RTI4017. 
MP acknowledges funding through the German Aerospace Center (DLR) under grant number 50OR2514.\\
The first author used ChatGPT (OpenAI) and Claude (Anthropic) for language editing and manuscript preparation. All scientific content and the final manuscript were reviewed and approved by the authors.
\end{acknowledgments}

\appendix

\section{Emission line fitting and BIC estimate}\label{sec:app_Hafit}
We fitted the SDSS spectrum of J1047 in a rest-frame window of $[-250,+250]$\,\AA\ around H$\alpha$, including also the [N\,\textsc{ii}]\,$\lambda\lambda6548,6584$ doublet. The fitting function models the continuum with a power law, which provides a good enough approximation given the limited wavelength range. The narrow components of \Ha and \NII  share the same redshift $z$ and FWHM (allowed to vary between 50 and 500\,km\,s$^{-1}$), and the \NII doublet ratio is fixed to 1:3. The broad H$\alpha$ component is described by three alternative models: (i) a single Gaussian, with FWHM$_{\rm BL}$ in the range 500--5000\,km\,s$^{-1}$ and redshift $z_{\rm BL}$ free within $\pm650$\,km\,s$^{-1}$ of the narrow-line redshift; (ii) two Gaussians with independent amplitudes and FWHMs but a common redshift $z_{\rm BL}$; (iii) the electron-scattering model of Sect.~\ref{sec:Hafit}, in which a Gaussian BLR profile $G_{\rm BLR}$ is mixed with its convolution with a symmetric exponential $E(\lambda;W)$, so that the broad profile is $f_{\rm scatt}\,E*G_{\rm BLR}+(1-f_{\rm scatt})\,G_{\rm BLR}$, with $0<f_{\rm scatt}<1$ and $W\in[500,5000]$\,km\,s$^{-1}$. The convolution is computed analytically, and the total broad-line flux (core plus scattered wings) is conserved. The posterior distributions were sampled with an MCMC routine (\texttt{emcee}; 256 walkers, $10^4$ steps) using the QubeSpec fitting package\footnote{\url{https://github.com/honzascholtz/Qubespec}}, with uniform or log-uniform priors on all parameters except $z$ and $z_{\rm BL}$, which have truncated Gaussian priors centered on the SDSS redshift. Best-fit values and uncertainties are the median and the 16th--84th percentile range of the posteriors.

To compare the three models, we used the Bayesian Information Criterion,
\begin{equation}\label{eq:BIC}
{\rm BIC}=\chi^2_{\rm min}+k\ln n ,\qquad
\chi^2_{\rm min}=\sum_{i=1}^{n}\frac{\left[F_i-M_i(\hat{\theta})\right]^2}{\sigma_i^2},
\end{equation}
where $F_i$ and $\sigma_i$ are the observed flux and its uncertainty in the $i$-th of $n$ spectral pixels inside the fitting window, $M_i(\hat{\theta})$ is the model evaluated at the best-fit parameters $\hat{\theta}$, and $k$ is the number of free parameters ($k=9, 11, 11 $ for the single-Gaussian, double-Gaussian and electron-scattering models, respectively). Both estimators in Eq.~\ref{eq:BIC} were computed over the same wavelength range for all models, and over the full wavelength range used for the fit, ensuring that the estimates include the full broad-wing profiles of the line. We adopt $|\Delta{\rm BIC}|>10$ as the threshold for strong evidence.  The electron-scattering model is preferred, with $\Delta{\rm BIC}=-552$ relative to the single-Gaussian model and $\Delta{\rm BIC}=-34$ relative to the double-Gaussian model (see Fig.~\ref{fig:line_profiles}).

\section{Comparison with other scattering medium configurations}\label{app:tauff_grid}
\begin{figure}[h!]
    \centering
    \includegraphics[width=1\linewidth]{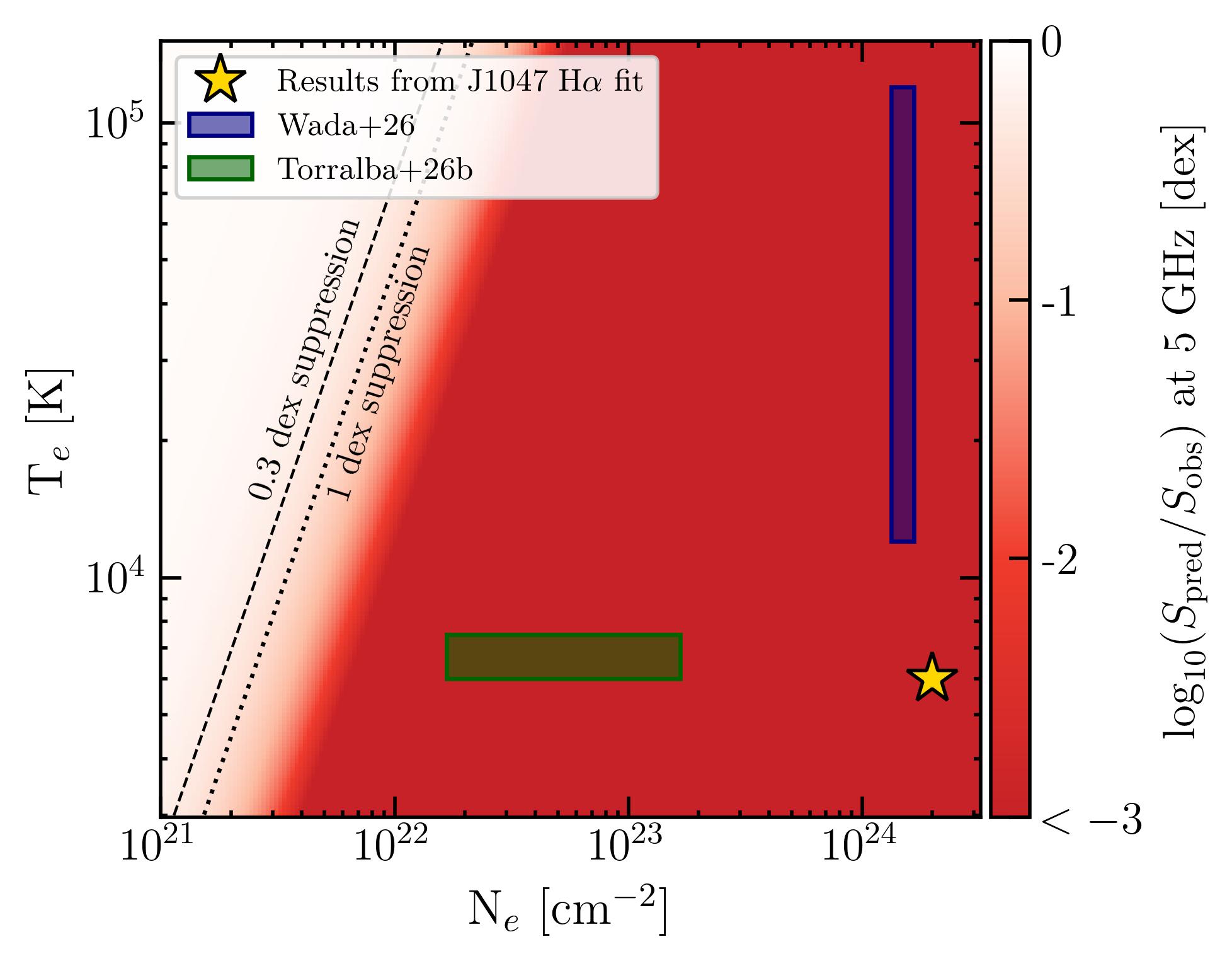}
    \caption{Deviation of the observed radio flux from the intrinsic spectrum at 5~GHz considering a grid of $\tau_{ff}$ computed using Eq.~\ref{eq:tau_ff_Ne} over $10^{21}<N_e/\rm cm^{-2}<3\times10^{24}$, and $3000<T_e/K<2\times10^5$ and fixing $z=0.1682$ and $r=0.01$pc. The result from the physical parameters inferred from the \Ha fit is reported with a golden star, while the range of flux suppression implied by assuming the physical properties of the scattering medium of \cite{Wada26} and \cite{Torralba25} are reported in blue and green, respectively. }
    \label{fig:abs_test}
\end{figure}

The $N_e$ and $T_e$ values that we got from Eq.~\ref{eq:Ne}, and Eq.~\ref{eq:Te} could be considered as "effective" values of the scattering gas physical properties, since they are obtained from the integrated line fit. Recent works suggested that the scattering medium, if present, might be in two possible configurations. \cite{Wada26}, using detailed CLOUDY modeling to study the Br$\gamma$ line profile of a local Type I AGN observed with GRAVITY \citep{GRAVITY17}, suggested that the broad wings might be produced by hot ionized gas very close to the central source, at $r\sim0.001$pc with $10^4<T_e/K<10^5$ and $0.8<\tau_e<1$. Instead, \cite{Torralba25}, analyzing the \Ha and \Hb broad line shapes of one high-z LRD suggested that the scattering might come from a warmer medium  $5000<T_e/K<7000$, outside the dense neutral gas shell, and with $0.01<\tau_e<0.1$. We tested these other possible configurations in Fig.~\ref{fig:abs_test}, by estimating the deviation from the intrinsic spectrum ($S_{\nu,int}\propto S_{5GHz}^{\alpha}$, with $\alpha=-0.85$ ) of the observed one ($S_{\nu,obs}= S_{\nu,int}\ e^{-\tau_{ff}}$), computing $\tau_{ff}$ over a wide $N_e$ and $T_e$ grid and fixing $z=0.1682$ and $r=0.01$pc. Neither of the alternative combinations of $T_e$ and $N_e$ proposed by \cite{Wada26} and \cite{Torralba25} would allow us to observe J1047 radio emission at 5~GHz. 

\section{SED fitting description}\label{sec:app_sed}
\begin{figure}[h!]
    \centering
    \includegraphics[width=1\linewidth]{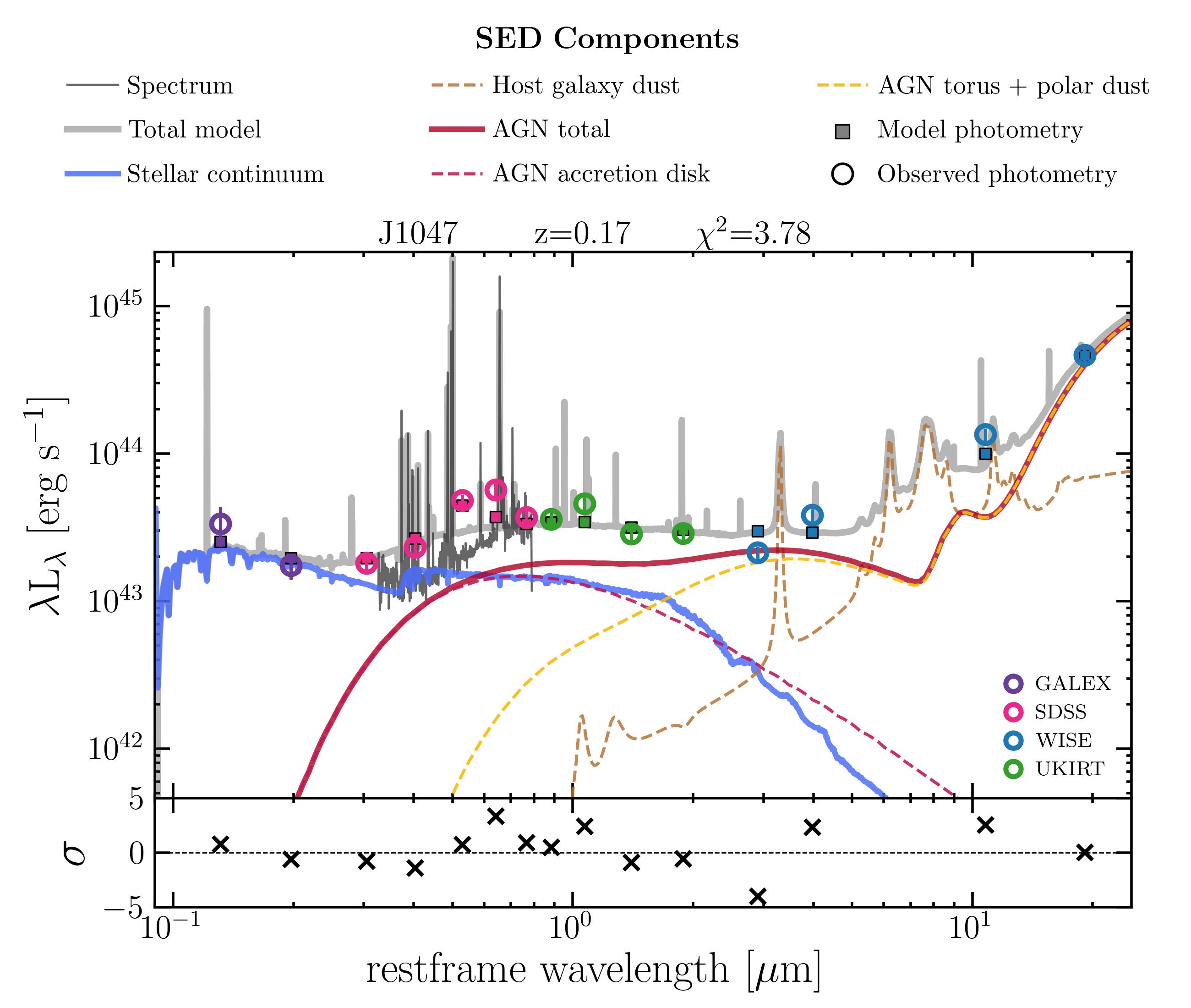}
    \caption{SED fitting decomposition of J1047. Observed data points and model predictions in the different filters are color-coded as labeled. The blue line represents the stellar contribution, while the red solid line represents the AGN contribution (sum of obscured disk emission, torus emission, and polar dust emission). The brown line represents the host galaxy dust emission. The final model is reported in light gray, while the SDSS spectrum of J1047 is in dark gray.}
    \label{fig:SEDfit}
\end{figure}
We collected the photometry of J1047 as done in \cite{Lin26_localLRD}: FUV and NUV from GALEX DR6; ugriz PSF photometry from SDSS DR17; grizy PSF photometry from Pan-STARRS DR2; $griz$ model photometry from Legacy Surveys DR10 (A. Dey et al. 2019); YJHK PSF photometry from UKIDSS DR11PLUS; and W1, W2, W3, W4 photometry from the Widefield
Infrared Survey Explorer (WISE). The SED fitting was performed using \texttt{CIGALE} \citep{boquien19,yang20}. 
We used delayed star-formation history (SFH) model, which can reproduce both early- and late-type galaxies, with an additional term that allows for a recent burst of star formation. We provide a range of ages of the main stellar population varying between 10 Gyr and 150~Myr, to account also for young stellar populations potentially dominating the rest-frame UV, and we allow the burst to happen within the last 100~Myr and to contribute up to 20\% of the total mass budget. We adopted stellar templates from \cite{bruzual03}, and a \cite{Chabrier03} initial mass function. We fixed the stellar and gas metallicity to that derived from the $T_e$ method by \cite{Izotov07} ($Z\sim 0.2Z_{\odot}$). For the attenuation of the stellar continuum emission, we use the \texttt{dustatt\_modified\_starburst} module with a Small Magellanic Cloud extinction law \citep{Pei92}. We select a narrow range of $E(B-V)_{\text{line}}$, the attenuation of the nebular emission lines, to match the result obtained from the spectral analysis ($A_V\sim 0.8$ from \citealt{Lin26_localLRD}) and then set two possible ratios between $E(B-V)_{\text{line}}/E(B-V)_{\text{star}}=[0.44,1]$. We also include dust emission in the infrared following the model by \cite{Draine14}. 

For the AGN contribution, we employed the \texttt{skirtor2016} \citep{Stalevski16} module updated by \cite{yang20}, that combines emission from the accretion disk, torus, and polar dust. The accretion disk, responsible for the UV-optical emission in the central region, is parametrized according to \cite{Schartmann05}, allowing for a deviation of the intrinsic slope in the $0.125<\lambda_{rest}/\mu m<10$ according to the parameter $\delta$ ($f_{\lambda}\propto\lambda^{-1.5\pm \delta}$). We allowed $-1<\delta<1$. 
Photons from the accretion disk can be obscured and scattered by dust in the vicinity, within the torus and/or in the polar direction. For the torus, the \texttt{skirtor2016} module employs a clumpy two-phase model \citep{Stalevski16}. For the polar dust component, we used the Small Magellanic Cloud (SMC) extinction curve and provided a range $0.05<E(B-V)<1.5$.  We allowed the opening angle of the AGN torus to vary between 40°$<\theta<$80° and the AGN line-of-sight inclination 20°$<i<$90° (where values 0°$<i<$90°-$\theta$ correspond to Type I AGN, while 90°-$\theta<i<$ 90° to Type II AGN). Finally, we performed two set of fits one in which we leave the AGN fraction (defined as the ratio between the AGN luminosity and the total galaxy luminosity between 0.63 and 0.65$\mu$m) to vary within 0.3 and 0.9, and the second in which we imposed an AGN fraction close to $\sim 85\%$ as inferred in \cite{Lin26_localLRD}. The best fit solution (shown in Fig.~\ref{fig:SEDfit}, $\chi^2_{red}=3.8$) is obtained with the flexible AGN fraction, which results in $f_{AGN}=0.41\pm0.03$, half the one estimated in \citealt{Lin26_localLRD}. Fixing the AGN fraction still returns a good fit, but with a worse $\chi^2_{red}=8.1$, and with a lower SFR as reported in Sect.~\ref{sec:SFR}


\section{Free-free absorption for the LOFAR non-detection}\label{app:LOFAR}
\begin{figure}[ht]
    \centering
    \includegraphics[width=1\linewidth]{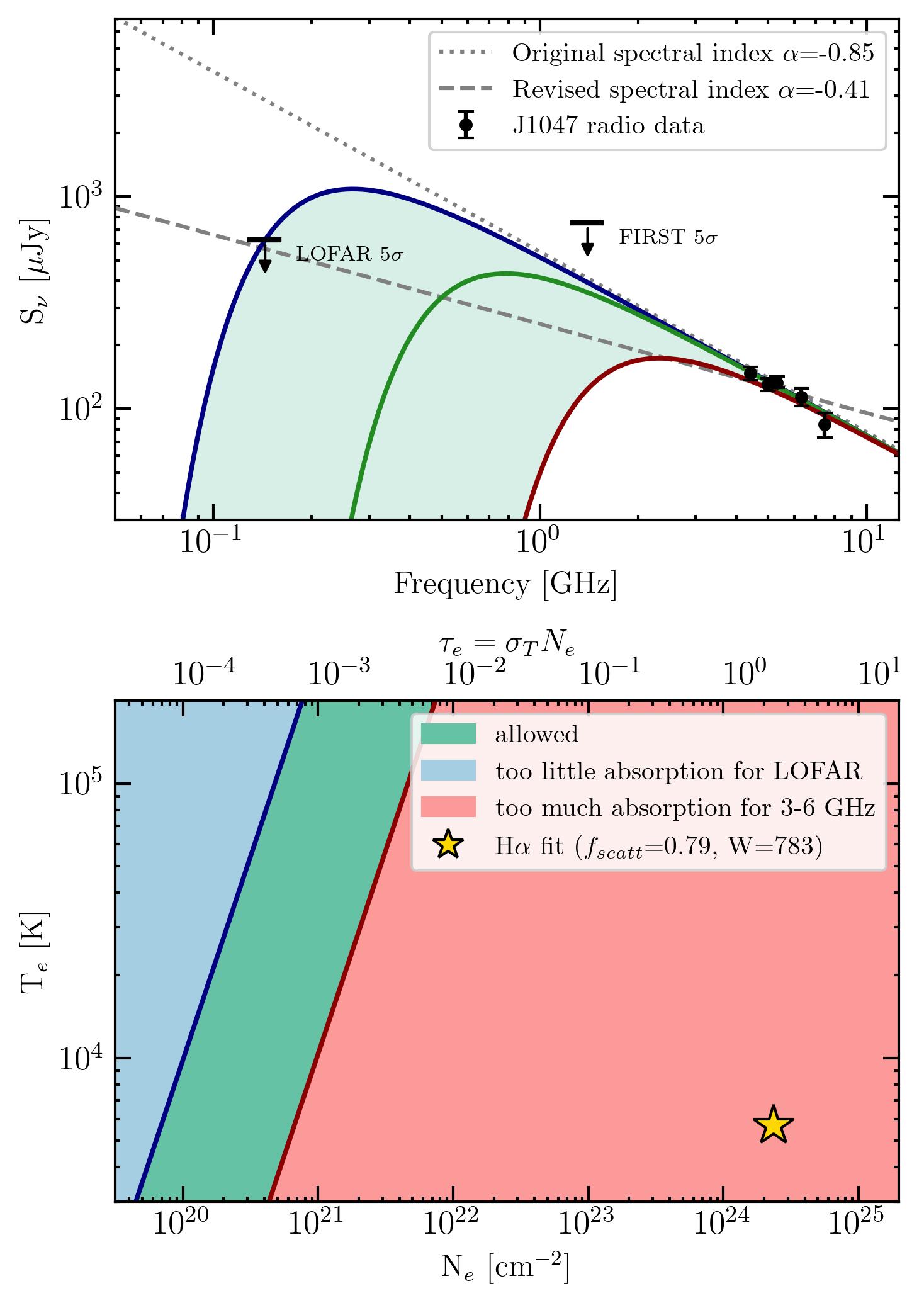}
    \caption{Upper panel: observed radio spectrum of J1047 compared to the original spectral shape derived in \cite{Rodriguez26_radioLRD} (gray dotted), the one consistent with the LOFAR upper limit (gray dashed, see Sect.~\ref{sec:LOFAR}), and the free-free attenuated maximal and minimal solutions as described in Appendix~\ref{app:LOFAR} (red solid and blue solid, respectively). Lower panel: allowed and forbidden combinations of $T_e$ and $N_e$ necessary to reproduce the observed spectrum of J1047 assuming that it is attenuated by FFA. The golden star represents the combination of parameters derived from the \Ha exponential fit in Sect.~\ref{sec:Hafit}.}
    \label{fig:LOFAR_Ne}
\end{figure}
We quantify the free-electron column density $N_e$ and corresponding Thomson optical depth $\tau_e$ required for a uniform ionized slab to reproduce the radio data of J1047 at both GHz frequencies and at 144 MHz. We assume an intrinsic power-law spectrum, $S_\nu=S_5(\nu/5~{\rm GHz})^\alpha$, with $\alpha=-0.85$ measured from the high-frequency data. The observed flux is therefore $S_\nu e^{-\tau_{\rm ff}(\nu)}$. Since $\tau_{\rm ff}\propto\nu^{-2.1}$ (see Eq.~\ref{eq:tau_ff_Ne}), the absorption at 144MHz is related to that at 5GHz by $\tau_{\rm ff, 144~\rm MHz}\simeq1.7\times10^3\tau_{5~GHz}$.

The VLA detections limit the amount of absorption: we fitted the high-frequency data points with the absorbed power law, and defined the upper limit on $\tau_5$ as the value at which the $\chi^2$ increases by $\Delta\chi^2=4$, i.e. $\sim2\sigma$ with respect to the unabsorbed case. This gives $\tau_5\lesssim8\times10^{-2}$. The LOFAR non-detection requires a suppression of at least a factor of 4.4, which requires $e^{-\tau_{\rm ff}(0.144)}\leq1/5$ and hence $\tau_5\gtrsim\ln5\,(5/0.144)^{-2.1}\simeq9\times10^{-4}$. Thus, the allowed range from the two constraints is $9\times10^{-4}\lesssim\tau_5\lesssim8\times10^{-2}$. Inverting Eq.~\ref{eq:tau_ff_Ne}, for a slab thickness of $r=0.01$ pc, this corresponds to $1\times10^{20}\lesssim N_e/{\rm cm}^{-2}\lesssim9\times10^{20}$ at $T_e=10^4$ K, or, using Eq.~\ref{eq:Ne}, to $7\times10^{-5}\lesssim\tau_e\lesssim7\times10^{-4}$ (see Fig.~\ref{fig:LOFAR_Ne}). From Eq.~\ref{eq:Ne}, this would also imply an almost negligible scattering fraction $f_{\rm scatt}\lesssim0.06\%$, not able to reproduce the broad \Ha wings. Consequently, even if the LOFAR non-detection is due to FFA, the conclusions drawn in Sect.~\ref{sec:radio_freefree} and Sect.~\ref{sec:discussion} remain valid.


\bibliography{literature}{}
\bibliographystyle{aasjournalv7}



\end{document}